\documentclass[preprint2]{aastex}
\usepackage{graphicx}
\usepackage{natbib}
\newcommand{\lya}{Ly$\alpha$\ }
\newcommand{\erfc}{\ensuremath{{\rm erfc}}}
\newcommand{\psec}{\ensuremath{\, {\rm s}^{-1}}} 
\newcommand{\yr}{\ensuremath{{\rm yr}}} 
\newcommand{\Myr}{\ensuremath{{\rm Myr}}} 
\newcommand{\cm}{\ensuremath{{\rm cm}}} 
\newcommand{\km}{\ensuremath{{\rm km}}} 
\newcommand{\kpc}{\ensuremath{{\rm kpc}}}
\newcommand{\Mpc}{\ensuremath{{\rm Mpc}}}
\newcommand{\K}{\ensuremath{{\rm K}}} 
\newcommand{\mK}{\ensuremath{{\rm mK}}} 
\newcommand{\erg}{\ensuremath{{\rm erg}}} 

\newcommand{\keV}{\ensuremath{\, {\rm keV}}}
\newcommand{\mev}{\ensuremath{{\rm MeV}}}
\newcommand{\Hz}{\ensuremath{\, {\rm Hz}}} 
\newcommand{\kHz}{\ensuremath{\, {\rm kHz}}} 
 
\newcommand{\sr}{\ensuremath{\, {\rm sr}}}

\newcommand{\Msun}{\ensuremath{{M_\sun}}}
\newcommand{\Lsun}{\ensuremath{{L_\sun}}}

\begin{document}

\title{The 21-cm Signature of the First Stars}
\author{Xuelei Chen}
\affil{National Astronomical Observatories, Chinese Academy of
  Sciences, Beijing 100012, China}
\email{xuelei@bao.ac.cn}
\author{Jordi Miralda-Escud\'e}
\affil{Institut de Ci\`encies de l'Espai (CSIC-IEEC)/ICREA,
  08193-Bellaterra, Spain \and}
\affil{Department of Astronomy, The Ohio State University, Columbus
  OH 43210, USA}
\email{miralda@ieec.uab.es}

\begin{abstract}

We predict the 21-cm signature of the first metal-free stars. The soft
X-rays emitted by these stars penetrate the atomic medium around their
host halos, generating \lya photons that couple the spin and kinetic
temperatures. These creates a region we call the {\it \lya sphere},
visible in 21-cm against the CMB, which is much larger than the HII
region produced by the same star. The spin and kinetic temperatures are
strongly coupled before the X-rays can substantially heat the medium,
implying that a 21-cm absorption signal from the adiabatically
cooled gas in Hubble expansion around the star is expected when the
medium has not been heated previously. A central region of emission from
the gas heated by the soft X-rays is also present although with a weaker
signal than the absorption. The \lya sphere is a universal signature
that should be observed around any first star illuminating its vicinity
for the first time. The 21-cm radial profile of the \lya sphere can be
calculated as a function of the luminosity, spectrum and age of the
star. For a star of a few hundred $\Msun$ and zero metallicity (as
expected for the first stars), the physical radius of the \lya sphere
can reach tens of kiloparsecs. The first metal-free stars should be
strongly clustered because of high cosmic biasing; this implies that the
regions producing a 21-cm absorption signal may contain more than one
star and will generally be irregular and not spherical, because of the
complex distribution of the gas. We discuss the feasiblity of detecting
these \lya spheres, which would be present at redshifts $z\sim 30$ in
the cold dark matter model. Their observation would represent a direct
proof of the detection of a first star.

\end{abstract}

\section{Introduction}

  The formation of the first stars marks the end of the cosmic dark
age and the beginning of the reionization era. The cold dark matter 
model with a cosmological constant ($\Lambda$CDM) of
structure formation predicts that
the first stars were formed in dark matter halos of $\sim 10^6 \Msun$,
which were sufficiently massive to allow the halo gas to collapse and 
cool through emission in molecular hydrogen roto-vibrational lines
(for reviews see, e.g., \citealt{BL01}; \citealt{M03}; \citealt{CF05}).
Numerical simulations and analytic arguments suggest that in the absence
of heavy elements, the gas in the halo center will cool hydrostatically
and collapse into a central object when the Jeans mass is $\sim 10^3
\Msun$, probably resulting in the formation of one or a few stars with
$\sim 100 \Msun$ \citep{BCL99,ABN00,OP03}. Stars of this mass live for
only $3 \times 10^6 \yr$ \citep{BKL01}. At the end of their lives the
supernova explosions enrich the intergalactic medium with heavy
elements, which accelerate the cooling process and allow less massive
stars to form. The reionization of the medium advances as more sources
of ionizing radiation (stars or black holes emitting light as quasars)
are formed in increasingly massive halos.

  A very powerful observational probe of the end of the dark age is
21-cm tomography, which provides information on the
state of the atomic intergalactic medium (IGM)(\citealt{MMR97,TMMR00}, 
for a recent review see e.g. \citealt{FOB06}). Hydrogen atoms interact
with 21-cm photons by radiative transitions between the hyperfine
split ground state. The hydrogen may be seen in emission or
absorption against the Cosmic Microwave Background (CMB) depending on
whether its spin temperature is above or below the CMB temperature. The
spin temperature is a weighted average of the gas kinetic
temperature and the CMB temperature: it is driven toward the CMB
temperature by the 21-cm transitions, and towards the kinetic
temperature by atomic collisions and by resonant scattering of \lya
photons \citep{W52,F59}. During the dark age (before the appearance of
sources of light such as stars or accreting black holes), 
the kinetic temperature drops below the CMB
temperature because of adiabatic cooling\footnote{This picture of dark
age would be drastically changed if non-conventional source of light,
  e.g. decaying dark matter, were present. 
For discussions on the state of IGM and 
21cm signal in such cases, see, e.g., \citet{CK04,FOP06,RMF07,VFMP07}.}, 
and collisional coupling of
the spin and kinetic temperatures can produce an absorption signal
\citep{LZ04,BL05,NB05}.
Collisional coupling becomes ineffective below a redshift $\sim 40$,
except in high density regions such as collapsed minihalos
\citep{I02,K06}.
At lower redshift, the gas may again appear in absorption when
illuminated with the \lya photons from the first sources, which also
couple the spin and kinetic temperatures through the Wouthuysen-Field
mechanism \citep{CM04,H06,PF06}.
At the same time, the ionizing photons from the same first
sources start heating the neutral medium, due mainly to soft X-rays
which penetrate deeper into the atomic regions \citep{CM04,S05}. The
21-cm fluctuations are then due to variations of neutral density, \lya
flux and spin temperature \citep{BL04}. As reionization proceeds, X-rays
heat the kinetic temperature to values much higher than the CMB
temperature, and at the same time a \lya photon background is produced
which strongly couples the spin and kinetic temperatures.
The 21-cm fluctuations are then essentially proportional to the density
of neutral hydrogen, dependent on the IGM density and the ionization
state \citep{MMR97,CM03,FSH04,K05}. 

  In the present work, we study the 21-cm signature of the first stars,
when the region around a star is illuminated by \lya photons for the
first time after the recombination epoch. Before a \lya photon
background is created, the spin temperature is close to that of the CMB
temperature at most spatial locations. However, near a first star,
the \lya photon flux is strong enough to change the spin temperature,
rendering the medium observable with 21-cm tomography. This region is
substantially greater than the HII region generated by the star and the
virial radius of the star-forming halo, as we shall see below. We shall
refer to such a region as a {\it Ly$\alpha$ sphere}. 
The gas in the inner part of a \lya sphere is heated up quickly by
the radiation of the star (which, owing to the surface temperature
$T_{eff}\simeq 10^5$ K of massive, metal-free stars, emits substantial
soft X-rays during the main-sequence lifetime) to a temperature above
$T_{CMB}$. Therefore, this inner region produces instead a core of 21-cm
emission. The heating rate is much lower at the outer part of the \lya
sphere. Depending on the spectrum and the age of the star, the gas in
the outer part of the \lya sphere should appear as a halo of 21-cm
absorption. This outer halo of absorption is not present if the region
has been illuminated before by a sufficiently strong background of soft
X-rays, and therefore heated by other more distant sources prior to the
appearance of the star. Therefore, the detection of the absorption
halo provides a unique signature of a first star.

  As this work was being completed, we became aware of the work by Cen
(2006) along similar lines. Our conclusions differ from those of Cen
in two respects: first, we find that \lya spheres of absorption would
be produced only around the very first sources of light in the universe,
which should be metal-free stars (see, e.g. \citealt{ABN02},
\citealt{BL04rev}). At later times, the medium should be rapidly heated
by X-rays and a global \lya background should be produced, eliminating
the \lya spheres. Second, the most important source of \lya photons that
couple the spin and kinetic temperatures around a first star are not the
continuum photons from the stellar photosphere, but the secondary \lya
photons due to impact excitations by electrons produced by the X-rays
themselves. As we shall see in this paper, this fact is essential for
improving the observability of these \lya spheres in 21-cm absorption,
by increasing their contrast relative to the background.

  In \S 2, we derive the properties of an individual \lya sphere. In \S
3 we investigate the formation history of the first stars. The growth of
the \lya and soft X-ray backgrounds, which affect the global evolution
of the kinetic and spin temperature of the IGM, is considered in \S 4.
In \S 5, we consider the observability of these \lya spheres. We
summarize our results in \S 6. In the following, we have adopted the
WMAP three year $\Lambda$CDM model, with $h=0.73$, $\Omega_m=0.23$,
$\Omega_b=0.047$, $n_s=0.99$, $\sigma_8 =0.77$.

\section{Properties of \lya spheres}

\subsection{Basic concepts and notation}

  The form of the \lya sphere produced around a first star, detectable
by means of the 21-cm emission and absorption, is determined by the
intensity of \lya photons and soft X-rays produced over the history of
the stellar lifetime. As a result of the irradiation of the neutral
medium by stellar ultraviolet light, \lya photons are produced by two
mechanisms: they may have been emitted directly from the stellar
photosphere at a frequency in the range between \lya and Ly$\beta$ and
then redshifted to \lya, or they may result from ionizing photons
which can ionize hydrogen atoms that subsequently recombine and also
excite other hydrogen atoms with the energetic electrons created in the
ionizations. Following \cite{CM04}, we designate the first type {\it
continuum} photons, since they enter the \lya line from the continuum
spectrum on the blue wing of the line, and the second type {\it
injected} photons, since they are injected close to the line center by
an atom that has recombined or has been excited by a secondary electron,
and has then reached the 2p state. We note that injected photons may
also originate from photons emitted between the Ly$\gamma$ line and the
Lyman limit by the star, which will be absorbed by a hydrogen as they
reach a Lyman resonant line, and may then produce a \lya photon if the
hydrogen atom cascades down to 2p. However, the majority of these
photons result in a cascade that ends in the 2s state instead, so these
typically make a small contribution to the creation of injected \lya
photons compared to the ionizing photons (see Hirata 2006; Pritchard \&
Furlanetto 2006).

  A fixed point in the intergalactic medium can be reached by continuum
photons from stars out to a maximum redshift of $z_{c,max}=(8/9)/(3/4) -
1 = 5/27$. Injected photons from ionization may come from a higher
redshift, but they need to arise from X-rays which can penetrate over a
large distance in the atomic medium. Every X-ray can produce several
secondary ionizations and collisional excitations by the electrons that
are produced, and these will result in the production of a \lya photon
when the hydrogen atom reaches the 2p (rather than the 2s) state. These
\lya photons start scattering after they are produced, and they are
confined to a very small region of the universe owing to the large cross
section for \lya scattering in the neutral medium. The flux of \lya
photons at any point in the atomic intergalactic medium is therefore
related to the local flux of X-rays and continuum photons between the
Lyman series lines coming from distant stars, although some spatial
diffusion of the \lya photons may take place in the last few
scatterings, when the photon exits the \lya line \citep{LR99}.

The brightness temperature of 21-cm photons is given by 
\begin{eqnarray}
\label{eq:dT}
\delta T_b &=& 40 \mK  \left( {\Omega_b h_0\over 0.03} \right)\,
\left( {0.3\over \Omega_{m0}} \right)^{1/2}
\left( {1+z\over 25} \right)^{1/2}\, \times \nonumber\\
&& \times {\rho_{HI}\over \bar\rho_H}\,
{ T_s - T_{CMB} \over T_s} ~.
\end{eqnarray}
The spin temperature $T_s$ is a
weighted average of the CMB temperature and the gas kinetic
temperature, and can be approximately calculated as (see Hirata 2006
for a more exact treatment)
\begin{equation}
\label{eq:Ts}
T_S = \frac{T_{CMB}+(y_{\alpha}+y_c) T_k}{1+y_{\alpha}+y_c}
\end{equation}
where the coupling coefficients $ y_{\alpha}$ and $y_c$  
are proportional to the  \lya scattering rate \citep{W52,F59} and 
spin-changing collision rate, respectively:
\begin{equation}
\label{eq:yalpha}
y_{\alpha} = \frac{P_{10}T_*}{A_{10}T_k}, \qquad 
y_c = \frac{C_{10}T_*}{A_{10}T_k}.
\label{eq:ya}
\end{equation}
Here, $T_* =0.068 $K is the temperature corresponding to the energy of
the hyperfine structure of hydrogen, $A_{10} = 2.87 \times 10^{-15}
\psec$ is the Einstein coefficient of the hyperfine structure levels,
and $C_{10}$ is the collisional de-excitation rate (tabulated first by
\citet{AD69}, and more recently by \citet{Z05,FF07a,FF07b}; in this
paper we use the tabulated values given by \citet{FOB06}). The spin-flip
transition rate $P_{10}$ is given by 
\begin{equation}
\label{eq:Palpha}
P_{10} = \frac{4}{27} H \tau_{GP}
\frac{S_c J_c + S_i J_i}{\tilde J_0}, 
\end{equation}
where $H$ is the Hubble rate at that redshift, $\tau_{GP}$ is the
Gunn-Peterson optical depth, $J_c$ and $J_i$ are the intensities of 
continuum and injected \lya photons, and $S_c$ and $S_i$ are numerical
factors of order unity that depend on the gas temperature and on
$\tau_{GP}$ (see \cite{CM04,H06}
\footnote{The numerical values of the heating and
collisional rate coefficients $I_i$, $I_c$, $S_i$,$S_c$ given in
\citet{CM04} are all too large by a factor of $\sqrt{\pi}$, 
because of an incorrect normalization used in
the original code (this does not affect the basic 
conclusions reached in \cite{CM04}).
We thank Chris Hirata for pointing out this error to us. Also, 
loss of \lya photons by the two photon decay
process could further reduce the effective collision rate
slightly, see \citet{H06} for further information.} ).
In the high temperature limit, $S_{c,i}\to 1$.
For simplicity, in the present paper we shall assume $S_c=S_i$, and
use the value as given by the fitting formula of \cite{FP06}.
The fiducial intensity $\tilde J_0$ corresponds to a photon
density of one photon per hydrogen atom per log frequency,  
\begin{equation}
\tilde J_0 = \frac{c n_H}{4\pi \nu_{\alpha}} ~,
\end{equation}
where $n_H$ is the number density of hydrogen and $\nu_{\alpha}$ is
the frequency of \lya.
In ordinary units, $\tilde J_0\, h\nu_{\alpha} =
2.69\times 10^{-24}\, \times
(1+z)^3 (\Omega_b h_0^2/0.02)\, \erg\, \cm^{-2}\,
{\rm s}^{-1} \sr^{-1} \Hz^{-1}$. 
At $z \sim 20-30$, $y_\alpha \sim 1$ when $J/\tilde J_0 \sim 0.02$.

\subsection{\lya sphere model}

  The \lya sphere of a first star is the region where the spin
temperature is substantially modified by the \lya photon flux arriving
from the star, producing an observable signature in the 21-cm line.
Stars that form later emit their
ultraviolet light into a region that has already been previously
illuminated by a background of light
from more distant stars which formed earlier, and the contrast of the
\lya sphere from an individual star will of course rapidly diminish as
the background intensity increases. In this paper, we designate as
``first star'' any star that emits the first light into the atomic
medium that surrounds it. Many more metal-free stars form later from
pristine material (i.e., not contaminated by any stellar ejecta), but
which has been illuminated by stellar light; these stars create a \lya
sphere in the presence of a light background, with lower contrast. Note
that the region around a star is also modified by an increased kinetic
temperature due to ionization by soft X-rays.

In this paper we assume that all the first stars are massive
\citep{BL04rev}, with
masses ranging from 25 \Msun ~ to 800 \Msun. 
The main sequence lifetime of a massive, metal-free star is about
three million years, and is generally much longer than the
Kelvin-Helmholz time to reach the main-sequence. We treat the stars as
static during their lifetime, with constant luminosity. As we shall see
below, due to the $r^{-2}$ decrease of the photon flux, the size of \lya
sphere is typically limited to a few tens of kpc, so the corresponding
light propagation time is also negligible compared with the stellar
lifetime. The scale of this \lya sphere is much greater than the halo
harboring the star, so we treat the density of the gas as
uniform and equal to the cosmic mean (note that this may be inaccurate
in many cases because of the highly biased large-scale distribution of
the first stars). The flux of the stellar continuum per unit frequency
is 
\begin{equation}
J_c = \frac{L(\nu_{\alpha})}{(4\pi)^2 r_p^2}
\end{equation}
where $r_p$ is the physical distance to the star, and $L(\nu)$ is the
stellar luminosity per unit frequency. 
We model the stellar radiation as blackbody with surface temperature
and total luminosity given by \citep{BKL01}
\begin{eqnarray}
T &=& 1.1 \times 10^5 \left(\frac{M}{100 \Msun}\right)^{0.025} K\\
L_{total} &=& 10^{4.5} \frac{M}{\Msun} \Lsun
\end{eqnarray}

\subsection{Ionization and heating of the \lya sphere}

We now proceed to solve for the ionization fraction, kinetic
temperature, spin temperature, and the 21-cm emission temperature
around the star. We include recombinations in our calculation, 
but as the recombination time at the average density of the universe
is a few times longer than the lifetime of the star and its
associated \lya sphere, it does not have a big impact on our result. 
Note that we are also assuming that
the high-density halo gas around the star (which does not have time to
be ejected from the halo during the stellar lifetime after it is ionized
and heated) does not absorb a large fraction of the stellar ionizing
photons emitted by the central star \citep{WAN04}. We also do not
consider ionization and
heating by any X-ray background that may have been produced by sources
other than the star. The neutral fraction is determined by
\begin{equation}
\frac{d x_{\rm HI}}{dt} = -  x_{\rm HI} \int_{\nu_H}^{\infty}
 d\nu \, F(\nu) \sigma_{\rm HI} (\nu) + \alpha_B^{\rm H}  
 x_{\rm HI} x_e n_{\rm H}
\end{equation}
\begin{equation}
\frac{d x_{\rm HeI}}{dt} = - x_{\rm HeI} 
\int_{\nu_{\rm HeI}}^{\infty} d\nu F(\nu) \sigma_{\rm HeI}
(\nu)+\alpha_B^{\rm HeI}  x_{\rm HeII} x_e n_{\rm He}
\end{equation}
\begin{eqnarray}
 \frac{d x_{\rm HeII}}{dt} &=& -\frac{d x_{\rm HeI}}{dt}-
x_{\rm HeII}  \int_{\nu_{\rm HeII}}^{\infty} d\nu F(\nu)
\sigma_{\rm HeII}(\nu) \nonumber\\
&&+ \alpha_B^{\rm HeII} x_e (1-x_{\rm HeI}-x_{\rm
  HeII}) n_{\rm He}
\end{eqnarray}
Here, $F(\nu)$ is the flux of ionizing photons per unit frequency, 
$n_{\rm H}$ and $n_{\rm He}$ are the total number density of H and He
atoms, and $x_i$, $\sigma_i(\nu)$, and $\nu_i$ are the fractional
abundance, the cross section and the ionization threshold frequency
of each atom or ion of type $i$, respectively. The case B
recombination rates for hydrogen $\alpha_B^{\rm H}$ is calculated by 
using the fitting formula given in \citet{HS63}, the 
$\alpha_B^{\rm HeI}$,  $\alpha_B^{\rm HeII}$ are calculated by 
interpolating/extrapolating the tabulated values given in \citet{SH95,HS98}.
In the helium
ionization, we have neglected the contribution of 
direct double ionization, as this rate is small compared with
ionization in two stages. 
We fit the cross sections using the functions given in \citet{V96}.
The flux at physical distance $r_p$ is given by 
\begin{equation}
F(\nu) = \frac{L(\nu)}{4\pi r_p^2} e^{-\tau(\nu)}
\end{equation}
where 
%\begin{eqnarray}
%\tau(\nu) &=& n_{\rm H} \sigma_{\rm HI}(\nu) \int_0^r dr' x_{\rm HI}(r')
%\nonumber\\
%&+&
% n_{\rm He} \sigma_{\rm HeI}(\nu) \int_0^r dr' x_{\rm HeI} (r') \nonumber\\
%&+& n_{\rm He}  \sigma_{\rm HeII}(\nu) \int_0^r dr' x_{\rm HeII} (r') 
%\end{eqnarray}
\begin{eqnarray}
\tau(\nu) &=& n_{\rm H} \sigma_{\rm HI}(\nu) \int_0^r dr' x_{\rm
  HI}(r')\nonumber\\
&&+n_{\rm He} \sigma_{\rm HeI}(\nu) \int_0^r dr' x_{\rm HeI} (r') \nonumber\\
&&+ n_{\rm He}  \sigma_{\rm HeII}(\nu) \int_0^r dr' x_{\rm HeII} (r') 
\end{eqnarray}

We assume $x_{\rm HI}=x_{\rm HeI}=2 \times 10^{-4}$, $ x_{\rm HeII}=0$
everywhere initially (to include the primordial ionization).
We then evolve the ion fractions on a 
lattice of 200 points (logarithmically distributed from $r=0.01 \kpc$
to $r=100 \kpc$) with the above differential equations. The
solution of the ordinary differential equation is obtained with 
a predictor-corrector scheme \citep{EU96}.

The internal energy of the gas is given by 
\begin{equation}
U=\frac{3}{2} \left[n_H (2-x_{\rm HI})+n_{\rm He} (3-2x_{\rm
    HeI}-x_{\rm HeII})\right] kT
\end{equation}
The volume heating rate $\Gamma=\frac{d U}{dt}$ is given by
\begin{equation}
\Gamma = \sum_{i={\rm HI,HeI,HeII}} n_i \int_{\nu_i}^{\infty} d\nu 
\, h(\nu-\nu_i) \eta(x,\nu) F(\nu) \sigma_i(\nu), 
\end{equation}
where the fraction of photo-electron energy converted to heat $\eta(x,\nu)$
is a function of energy and free electron fraction. \citet{SvS85}
calculated this function using a Monte Carlo simulation, and found
that it can be fitted as
\begin{equation}
\eta = 0.9971 (1-(1-x^{0.2663})^{1.3163}).
\end{equation}
Note that this fit breaks down at very low $x$, but it is a
good approximation for $x \gtrsim 10^{-4}$. Also, this fitting
function was obtained for non-primordial helium abundance, so the real
result may differ slightly.

At large $r_p$, where the ionized fraction in the \lya sphere is
low, cooling is negligible and we simply calculate the temperature
using the above expressions for calculating the internal energy and the heating
rate, assuming that there is no cooling. The effect of atomic cooling sets in
rather abruptly once the temperature reaches $T\simeq 10^4 $ K, and we
assume that the temperature stays fixed at this value after it is
reached. This does not affect our result on 21-cm because at these high
temperatures the emission depends only on the hydrogen column density.

\subsection{Secondary \lya photons from ionization by X-rays}

  Owing to the high temperature of the metal-free stars, the soft X-rays
from the stellar photosphere result in a substantial ionization and
heating ahead of the thin shell where most of the ionization occurs.
These high-energy electrons produced by the X-rays that penetrate into
the mostly atomic zone also result in the emission of additional \lya
photons: in the limit of low fractional ionization and high photon
energy, a fraction $\eta_{\alpha} \sim$ 40\% of the X-ray energy is
converted to photons through excitation by the high-energy electrons. 
Here we assume all of these are in \lya photons. In reality, the fraction
may be slightly smaller.The intensity of these
\lya photons is added to those coming from the stellar photosphere and
can contribute to the coupling of the spin and kinetic temperatures.

  The \lya photons produced through X-ray ionization are ``injected''
photons, since they are introduced at the resonance line center. Spatial
diffusion of these photons is negligible, because the width of the
scattering line is much less than the Hubble expansion velocity of the
\lya sphere at the radius where the emission of these photons is
important. To evaluate their intensity, we note that each X-ray will
produce a number of \lya photons $\eta_\alpha \nu/\nu_{\alpha}$, and
that each \lya photon produced in this way will have its frequency
redshifted at a rate $H\nu_\alpha$ if the effect of scatterings are
neglected, where $H$ is the Hubble constant. Hence, the number density
of these \lya photons per unit of frequency should be equal to the
heating rate per unit of volume (eq.\ 15) times the factor
$(\eta_{\alpha}/\eta)/(h\nu_{\alpha}) / (H \nu_{\alpha})$. Their
intensity is therefore equal to
\begin{equation}
J_i = { c \eta_{\alpha} \Gamma \over 4\pi h H \nu_{\alpha}^2 } ~.
\end{equation}
In terms of the fiducial intensity in eq.\ (5), we can rewrite this as
\begin{equation}
{J_i \over \tilde J_0} = { \eta_{\alpha} \Gamma \over
\eta h \nu_{\alpha} n_H H } ~.
\label{iint}
\end{equation}
Noting that the quantity $h\nu_{\alpha} n_H H$ is roughly the heating
rate required to heat the gas to a temperature $h\nu_{\alpha}/k_B\sim
10^5$ K over a Hubble time, we can easily see a relation between the
temperature to which the gas is heated by X-rays and the \lya intensity
that can be produced from the same X-rays. For example, considering the
gas temperature and \lya intensity at half the lifetime of the star,
$t=1.5\times 10^6 \yr \simeq 10^{-2} H^{-1}$, at a distant point in the
\lya sphere where the temperature has been heated due to X-ray
absorption by $\Delta T = 10$ K (still allowing for a large 21-cm
absorption signal), and using $\eta_{\alpha}/\eta\simeq 4$ (valid for
high neutral fractions and X-ray energies), an intensity $J_i/\tilde J_0
\simeq 4\times 10^{-2}$ can be generated, sufficient to obtain
$y_{\alpha}\simeq 2$ for the spin-kinetic temperature coupling.

  We note here that the scattering of \lya photons can also change the
kinetic temperature of the gas. As shown by Chen \& Miralda-Escud\'e (2004),
the heating rate due to the \lya scattering is negligible compared to
that caused by X-rays, and we do not include it here. In fact, for the
regime that we shall be interested here, the temperature of the atomic
medium is high enough that the scattering of \lya photons actually
causes {\it cooling} of the gas (see Fig.\ 3 in Chen \& Miralda-Escud\'e
2004), although by a negligible amount.

\subsection{Results on the ionization and temperature profile evolution}

\begin{figure}[htbp]
\begin{center}
\includegraphics[width=0.45\textwidth]{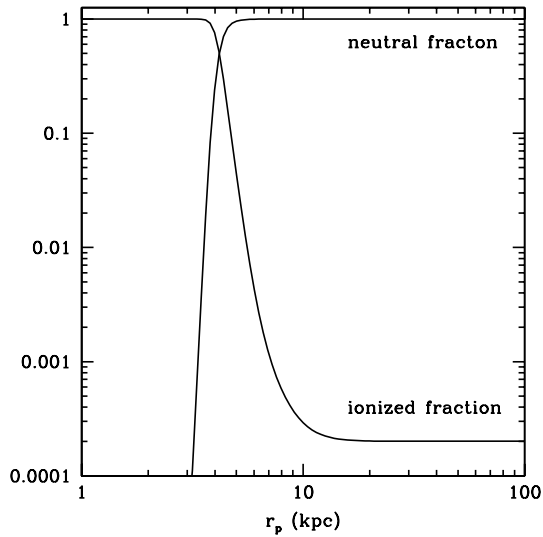}
\caption{\label{fig:singleion} This shows the neutral and ionizatgion 
fraction of H for a $200 \Msun$ star at $z=20$, with age 1.5 Myr.} 
\end{center}
\end{figure}

\begin{figure}[htbp]
\begin{center}
\includegraphics[width=0.45\textwidth]{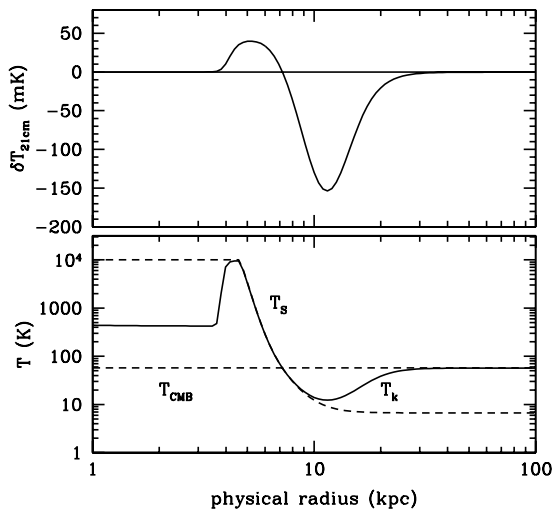}
\caption{\label{fig:sprofO} The kinetic and spin temperature (low
  panel) and 21-cm brightness temperature (upper panel) of a first star
  (same condition as above).}
\end{center}
\end{figure}

We plot the neutral and ionization fraction of the gas as a function
of distance from the star in Fig.~\ref{fig:singleion}, for a star of 
$200 \Msun$ at $z=20$ and age 150 Myr. As can be seen from the figure,
outside the HII region, the ionization fraction drops, but there is
still some residue ionization, which is produced by the penerating
X-ray photons. At about 15 kpc, the ionization fraction approaches
the background value. The spin and kinetic temperature profile, and
the 21-cm brightness temperature is plotted in Fig.~\ref{fig:sprofO}.

  The spin temperature is strongly coupled to the kinetic temperature
throughout the zone of the strong temperature drop ahead of the
ionization front. The injected \lya photons generated from the
ionization by X-rays are far more intense than the continuum \lya
photons originating from the stellar photosphere. The $y_{\alpha}$
parameter (see eq.\ \ref{eq:ya}), which indicates how strongly the spin
and kinetic temperatures are linked by the \lya photons, becomes of
order unity just at the radius where the atomic medium starts being
heated above its initial temperature by the X-rays from the star (for
the reason explained above in \S 2.4). At larger radius, the spin
temperature gradually rises above the kinetic temperature towards the
CMB temperature as the \lya coupling becomes weaker. This produces a
region of strong 21-cm absorption against the CMB, coming from the
atomic medium that is starting to be ionized and heated, still far from
the ionization front where most of the ionization occurs. The
temperature in this region has not yet increased very much, allowing for
a strong absorption signal, but the \lya photons generated by the
ionizations are already strong enough to lower the spin temperature by a
large factor below the CMB temperature. 
Note that at very large radius, the spin
temperature reaches a constant that is slightly below the CMB
temperature because of collisional coupling, causing a constant level
of absorption on the CMB.

\begin{figure}[htbp]
\begin{center}
\includegraphics[width=0.45\textwidth]{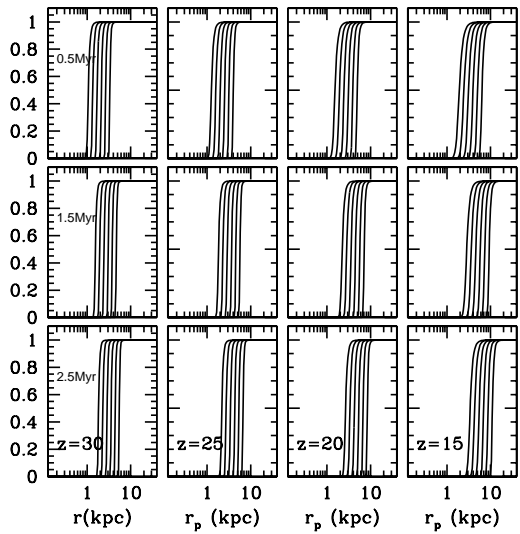}
\caption{\label{fig:ionprof} Neutral fraction as a function of
radius in the HII region at the indicated redshifts and times after the
birth of the star. In each panel, the six curves are for a star of mass
$M=25$, 50, 100, 200, 400, and 800 $\Msun$, from bottom to top.
}
\end{center}
\end{figure}

\begin{figure}[tb]
\begin{center}
\includegraphics[width=0.45\textwidth]{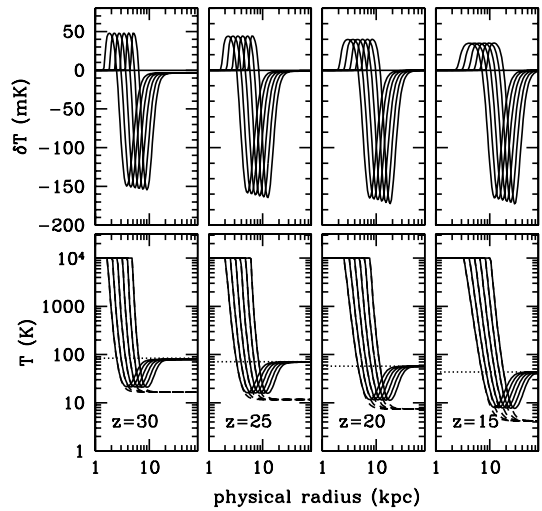}
\caption{\label{fig:profO} Brightness and temperature profiles of
\lya spheres, at a time $t=1.5$ Myr and the indicated redshifts.
{\it Bottom panels:} $T_k$ (dash lines), $T_{CMB}$ (dotted line), and
$T_S$ (solid lines). The six curves are for a
star mass $M=25$, 50, 100, 200, 400, and 800 $\Msun$. {\it Top panel}:
21-cm brightness temperature fluctuation on the CMB at time $t=1.5$ Myr.
}
\end{center}
\end{figure}

  Inside the HII region (the region that is mostly ionized), the kinetic
temperature is assumed to stay constant. The spin temperature should
also stay high, not only because of the \lya photons produced by the
high-energy electrons generated in the ionization front, but also
because of additional \lya photons generated by recombinations in the
HII region. We have not calculated the detailed shape of the spin
temperature profile in the HII region (this would be affected by
diffusion of \lya photons generated at the ionization front). However,
this profile does not actually matter very much because, in the regime
where the spin temperature is much higher than the CMB temperature, the
observed 21-cm brightness depends only on the column density of hydrogen
per unit of velocity, and not its spin temperature.

To see how these depends on the redshift, age, and mass of the star,
we made plots of Fig.~\ref{fig:ionprof} and Fig.~\ref{fig:profO}. 
In Fig.~\ref{fig:ionprof}, the top, middle, and bottom panels
correspond to a time after the birth of the star of 0.5, 1.5, and 2.5
Myr. The result is shown at redshifts $z=30, 25, 20, 15$ from left to
right. In each panel we plot the neutral fraction for six stellar
masses: 25, 50, 100, 200, 400, and 800 $\Msun$. The ionized regions grow
with the age of the star; for fixed stellar mass and age, the HII region
is smaller at higher redshift because the gas is denser.
The maximum physical radius reached is 1 to 10 kpc.
The kinetic and spin temperatures of the \lya sphere are plotted as a
function of physical distance in the lower panels of
Fig.~\ref{fig:profO}, at a time $t= 1.5$ Myr and the four redshifts
$z=30$, 25, 20, and 15. The six curves in each panel are for stellar
masses $M=25$, 50, 100, 200, 400, and 800 $\Msun$. We have assumed that
there is no global \lya background or heating prior to the formation of
the star. The kinetic temperature is equal to our imposed floor
($10^4$ K) within a few kiloparsecs, due to heating by ionization.
Beyond this radius the kinetic temperature drops rapidly as the atomic
zone is penetrated, until it reaches the constant value of the unheated
atomic medium far from the star, which is below the CMB temperature
owing to adiabatic cooling.

\begin{figure}[htbp]
\begin{center}
%\plotone{profO_noX.eps}
\includegraphics[width=0.45\textwidth]{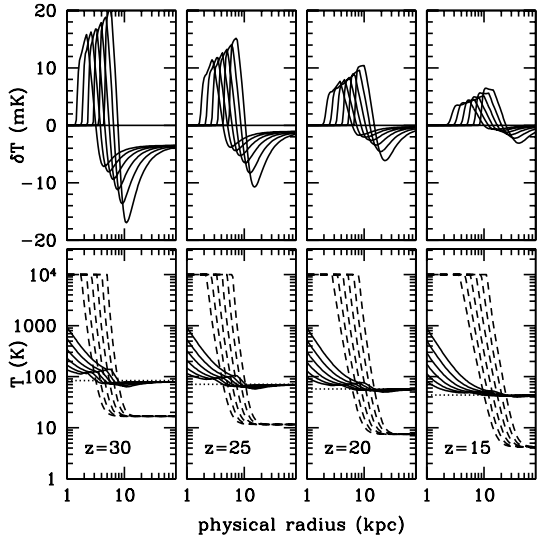}
\caption{\label{fig:profO_noX} The temperature profile without X-ray
  induced photons.}
\end{center}
\end{figure}

\begin{figure}[htbp]
\begin{center}
%\plotone{profO20.eps}
\includegraphics[width=0.45\textwidth]{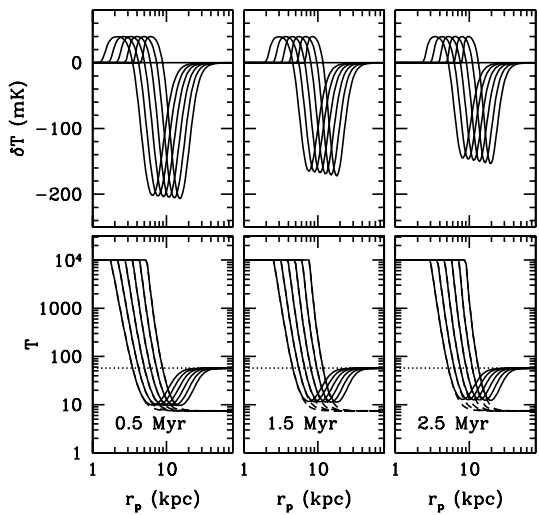}
\caption{\label{fig:profO20} Brightness profile of \lya spheres at
$z=20$, for the three indicated ages after the star birth.}
\end{center}
\end{figure}

In Fig.~\ref{fig:profO_noX}, we plot the temperature profile if the \lya
photons are produced only from the star, and not by X-ray
excitation. As can be seen from the figure, the coupling is much
weaker so the spin temperature is not strongly coupled to the kinetic
temperature, and the absorption signal is much weaker.

The \lya sphere evolves as the gas is heated and ionized by the star.
In Fig.~\ref{fig:profO20}, the same profiles are plotted at redshift
$z=20$ and three different ages, $t=0.5$, 1.5, and 2.5 Myr. The
features of the \lya sphere shift with time to larger radius and become
shallower as the region where ionization and heating are important
grows. A cross section map of the \lya sphere for a star of $200 \Msun$, age 1.5
Myr at $z=20$ with no global heating is shown in 
the left panel of Fig.~\ref{fig:cmap}. The central HII region where
$\delta T_b \simeq 0$, the emission region (red) and absorption region
(blue) is clearly seen. In real observations,  the absorption signal 
against the CMB at large radius is the dominant
signature of the \lya sphere formed around a first star: it is of much
higher brightness temperature and occupies a much larger area than the
emission region close to the star that has been heated by the
ionization. This small emission region could only be detected with
very high angular and frequency resolution. In the absence of
sufficient frequency resolution, the projected region of absorption on
top of the small emission region would greatly overwhelm the signal.

\section{Formation History}

  We now consider the formation history of the first stars in order to
estimate the abundance of the \lya spheres discussed in the previous
section. We calculate the comoving density of collapsed dark matter
halos with the halo model of Press-Schechter \citep{PS74}, and also the
Sheth-Tormen \citep{ST99} prescription. The density of collapsed halos
of mass $m$ at redshift $z$ can be written as
\begin{equation}
N(m,z) = \frac{\rho}{m} s f(s) \frac{ds}{dm}
\end{equation}
where $s=\delta_c^2(z)/\sigma^2(M)$, $\sigma(M)$ is the variance of the
density fluctuation smoothed on a mass scale $M$ with a top hat filter,
and the critical linear overdensity is
$\delta_c(z)=1.686/D(z)$, where $D(z)$ is the linear growth
factor. The function $f(s)$ is
\begin{equation}
s f(s) = A (1+\frac{1}{s'^p})\sqrt{\frac{s'}{2\pi}} e^{-s'/2}.
\end{equation}
Here, $s'=a s$, with $A=1/2, a=1, p=0$ for the Press-Schechter (PS)
mass function, and $A=0.322$, $a=0.707$, $p=0.3$ for the 
Sheth-Tormen (ST) mass functions.

  The gas in a dark matter halo can form a first star provided that
the primordial gas in the halo is able to cool by means of the small
fraction of molecular hydrogen that can be made at this early epoch,
faster than the rate at which the gas is dynamically shock-heated
during mergers and accretion (e.g., \cite{Y03}, \cite{R05}). We
assume for simplicity that star formation proceeds in any halo with 
$T_{\rm vir} > 2000 \K$, where $T_{vir}$ is the gas temperature after
virialization and in hydrostatic equilibrium. For neutral gas, the mass
of the halo when the virial temperature reaches this value is  
\begin{eqnarray}
M_{min}(z)&=& 1.17 \times 10^6 
\left(\frac{\Omega_m h^2}{0.147}\right)^{-1/2}\nonumber\\
&&\times \left[\frac{(1+z)}{21}\right]^{-3/2} \frac{T_{vir}}{2000 \K} ~.
\end{eqnarray}

  The lifetime of one of these massive, first stars is $t_* \approx 3
\Myr$, so only halos that have formed within the last $t_*$
would possess an active star. We assume here that only one metal-free star forms
per halo, since the ionizing radiation from these massive stars is able
to ionize and expel all the gas in the halo (e.g., \cite{WAN04}),
and any future stars will already form from metal-enriched gas
(presumably with a mass function closer to the present one). 

After its main sequence life time, a first star may explode 
as a supernova and destroy itself completely; 
or explode as a supernova and left behind a black hole or neutron
star; or collapse into a black hole directly \citep{WHW02}. In the
absence of continued supply of X-ray and \lya photons, the \lya sphere
dims as the scattering \lya photons diffuse out. However, supernovae remnant
and accreting black hole may become new source of light, the \lya
sphere could be maintained, or even grow to much larger size.
We will investigate these interesting possibilities in subsequent work.

For simplicity, we assume that one star is present in each halo that has
just formed with mass $M>M_{min}(z)$. At high redshift, when the fraction
of matter that has collapsed in halos in which gas can cool is small, the
number of halos of mass $m$ that have formed over a time $t_*$ can be
approximated as
\begin{equation}
\label{eq:nstar}
n_* = \int_z^{z_a} dz \int_{M_{min}(z)}^{\infty} dm ~
\frac{\partial N(m,z)}{\partial z} ~,
\end{equation} 
where the cosmic time at $z_a$ is $t(z_a) = t(z) - t_*$.
In the redshift range $15 \lesssim z \lesssim 40$, this can be
simplified as 
\begin{equation}
n_* \approx (1+z)H(z)t_*  \int_{M_{min}(z)}^{\infty} dm 
~\frac{\partial N(m,z)}{\partial z}
\end{equation}
At low redshift, the approximation obviously fails as halo destruction
by mergers becomes important. At high redshift $n(m,z)$ evolves too
rapidly and the integration over redshift needs to be done more
carefully.

\begin{figure}[tb]
\begin{center}
\includegraphics[width=0.45\textwidth]{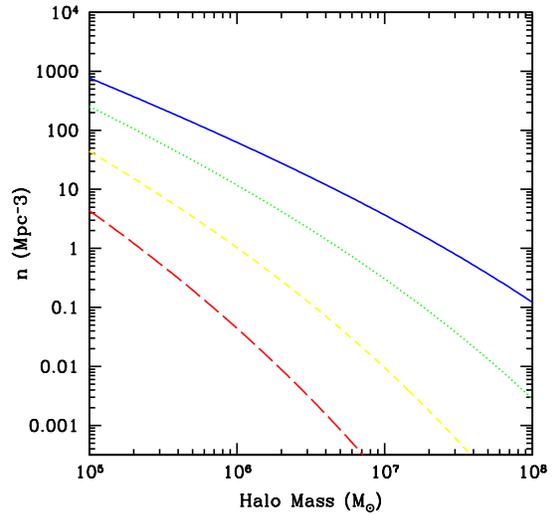}
\caption{\label{fig:halodist} Number density of star-harboring halos of
mass $m$ per comoving $\Mpc^{-3}$ per $\log m$.
The four curves are at $z=15$ (solid, blue
curve),$z=20$ (dotted green curve), $z=25$ (short-dash yellow curve),
and $z=30$ (long-dash red curve).}
\end{center}
\end{figure}

\begin{figure}[tb]
\begin{center}
\includegraphics[width=0.42\textwidth]{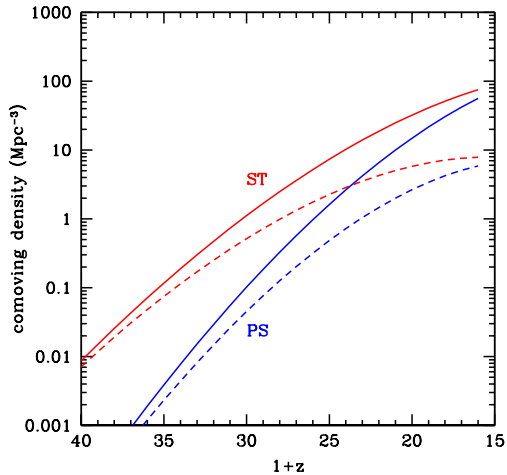}
\caption{\label{fig:haloden} Comoving density of halos and first
  stars. {\it Solid curves:} total density of star-forming halos.
  {\it Dashed curves:} density of star-forming halos formed in the
past 3 Myr. Results are shown for the Press-Schechter (PS) and
  Sheth-Tormen (ST) halo mass functions.}
\end{center}
\end{figure}

  We plot the mass function of halos formed within a time $t_*$,
calculated with the Press-Schechter formalism in Figure
\ref{fig:halodist}. In Figure \ref{fig:haloden}, the comoving densities
of star forming halos (solid curves) and first stars (dashed curve; this
is equal to the density of halos formed within the past three million
years) are shown, both for the Press-Schechter and Sheth-Tormen mass
functions. The ST mass function predicts a higher abundance for rare
fluctuations, so at very high redshift the ST result is substantially
greater than the PS result. At lower redshifts the ST
function fits N-body simulation results better, although it tends to
overpredict the number of halos at $z>10$ \citep{R03}. 
The accuracy of the mass functions at such high redshifts is still 
largely unknown, different conclusions
were reached in different numerical studies \citep{JH01,J01,R03,I06,Z06,W06,L07},
and in fact may also depend on the definition of the halo and its mass
in the simulation\citep{CW07}.

\begin{figure}[htbp]
\begin{center}
\includegraphics[width=0.4\textwidth]{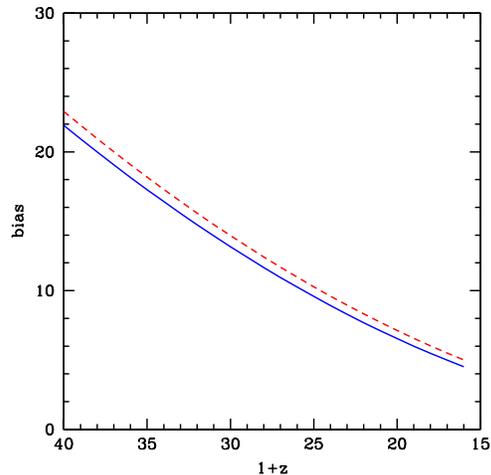}
\caption{\label{fig:bias} Bias factor of star-forming
  halos for the PS function (solid curve) and ST function (dashed curve).
}
\end{center}
\end{figure}

  The comoving density of first stars peaks around a redshift
$z \sim 15$. Our calculation begins to fail at this redshift for several
reasons: the merger destruction rate of the halos becomes comparable to
the formation rate, and equation \ref{eq:nstar} is no longer accurate.
Physical feedback effects suppressing the formation of metal-free stars
and new star formation in regions containing metals may become important
even earlier than $z=15$. We do not consider these uncertainties in this
paper, where we merely want to illustrate how the \lya spheres of the
first stars may interact, so our calculation is stopped at $z = 15$. 

\begin{figure}[htbp]
\begin{center}
\includegraphics[width=0.42\textwidth]{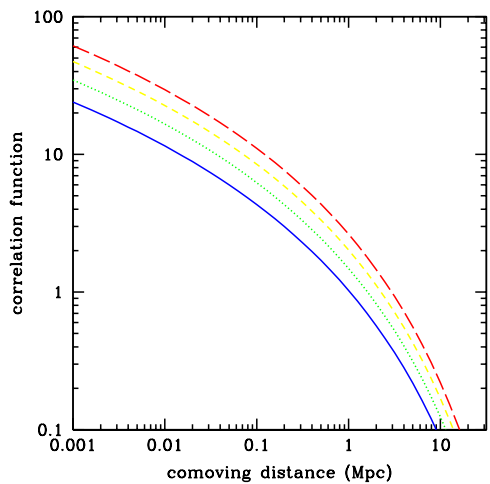}
\caption{\label{fig:corr} Correlation function of star-forming
  halos, at redshifts $z=15$, 20, 25, and 30 (from bottom up), calculated
  with the Press-Schechter function. Note that the correlation function
  increases with redshift, because of the increasing bias factor.}
\end{center}
\end{figure}

\begin{figure}[htbp]
\begin{center}
\includegraphics[width=0.42\textwidth]{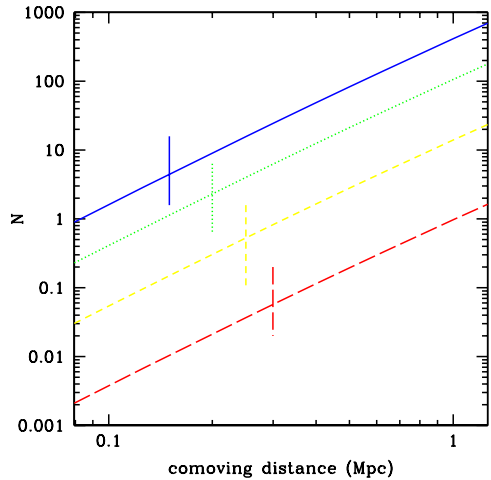}
\caption{\label{fig:cumu} Expected number of star-forming halos as
a function of distance. The four curves are 
 $z=15$ (solid, blue curve), $z=20$ (dotted green curve), 
$z=25$ (short-dash yellow curve),
and $z=30$ (long-dash red curve). A physical distance of 10 kpc
(the approximate size of the \lya spheres) is marked at each redshift.
}
\end{center}
\end{figure}

  Halos harboring the first stars are strongly clustered because of the
flatness of the Cold Dark Matter power spectrum at small scales, which
implies a high bias of the first objects in large-scale regions of
high density. The bias of halos of mass $M$ at redshift $z$ can be
approximated as 
\citep{MW96,MW02}
\begin{equation}
b=1+\frac{s(M,z)-1}{\delta_c(0)}
\end{equation}
for the PS function, and 
\begin{equation}
\label{eq:bias}
b=1+\frac{1}{\delta_c(0)} \left[s'+0.5 s'^{0.4}
-\frac{s'^{0.6}/0.84}{s'^{0.6}+0.14}\right]
\end{equation}
for the ST function. We calculate the bias corresponding to the minimal
mass for halos harboring stars at each redshift. The result is shown in
Fig.~\ref{fig:bias}. The bias grows very rapidly with redshift, faster
than $(1+z)$, implying that (perhaps counter-intuitively) the first stars
are increasingly clustered as the redshift increases, at a fixed
comoving scale.

  The correlation function of biased halos is given by 
\begin{equation}
\xi_{hh}(r,z) \approx b^2 \xi_{lin} (r,z) ~,
\end{equation}
where the linear correlation function is
\begin{equation}
\xi (r,z) = \int \frac{d^3 k}{(2\pi)^3} P_{lin}(k,z) e^{i kr} ~.
\end{equation}
We plot the correlation functions in Fig.~\ref{fig:corr}. The increasing
amplitude of the correlation function with redshift is again a result of
increasing bias factor with redshift.

  The average number of neighboring \lya spheres within a given radius
$r$ is (neglecting higher order clustering terms)
\begin{equation}
\bar{N}(R) = \bar{n}(z) \int_0^R d^3r \left[1+\xi(r)\right],
\end{equation}
where $\bar{n}(z)$ is the number density of star forming halos.
Note that this calculation is not very accurate, 
as non-linear effect may be important at such small scales, it is only
meant to give an idea on how the \lya spheres are correlated. 
The result is shown in Fig.~\ref{fig:cumu}. The average number of halos within
a sphere of physical radius $r=10$ kpc is $\sim 0.04$ at $z=30$ and
$\sim 4$ at $z=15$. This number increases with time because of the
growing number of collapsed halos, despite the decrease in the
correlation function. Even if several similar halos are 
found within the radius of the \lya sphere, the formation time of the
two stars may not coincide, thus the observed density of \lya sphere
is $f_* \bar{N}$, where the duty cycle factor $f_* \sim
n_*/n_{halo} < 1$. Moreover, feedback effects (e.g., from
molecular photodissociation) of the radiation of one star may
adversely affect the formation of the other one. 

  The large biasing factor of the star-harboring halos implies that the
first stars which produced \lya spheres may have an irregular
neighborhood, with more collapsed structures on the scale of the \lya
sphere than one would find in a random location of the universe. These
density fluctuations will likely make the ``\lya spheres'' have a
structure that is not actually very spherical, but irregular.

\section{The impact of \lya and hard X-ray backgrounds}

\subsection{The \lya background}

  In \S 2, we have computed the kinetic and spin temperature profiles of
the atomic medium around an isolated first star. The largest signal this
``\lya sphere'' produces in 21-cm is the absorption against the CMB at
large distances from the star. This absorption can be detected against
the background if the spin temperature of the atomic medium is coupled
to the kinetic temperature only close to the star, but not far from it
because of the absence of \lya photons; each star would then produce an
isolated \lya sphere. However, as the number of first stars increases, a
\lya background will be generated which can couple the spin and kinetic
temperatures everywhere, producing a global absorption signal against
which the regions near one star would be harder to detect. In fact,
owing to the presence of the foreground, it is not possible to make
absolute measurements of the 21-cm signal, but only its spatial 
variations and non-smooth variation in redshift. 
In Figures \ref{fig:profO} and 
\ref{fig:cmap}, the plotted
emission (absorption) temperature is not the absolute value, but the
difference its value far from any \lya sphere.

  The \lya background originates from the continuum photons
emitted by the stellar photospheres. This is because 
the injected photons resulting from
the high-energy electrons generated by soft X-ray photons are emitted at the \lya
line, and are rapidly redshifted as they leave the \lya sphere where
they have been created. On the other hand, hard X-ray photons may
penetrate the \lya
sphere and produce some injected photons very far from the sources, 
but the total energy released in hard X-ray photons by the
first stars is less than that in UV photons, and only a very small fraction of the
hard X-ray photon is absorbed in the IGM to produce 
injected \lya photons. Because the injected photons vastly dominate
the intensity over the continuum photons from the central star, the
background of \lya photons is much fainter than it would be if
continuum photons were the dominant ones. This allows for a lasting
contrast of individual \lya spheres in comparison to the background,
because the \lya background intensity stays low even when a very large
number of first stars are seen from any random point in the universe. 

  At the same time, we need to take into account the background of
X-rays produced by the stars and their global heating effect, which
will raise the kinetic and spin temperatures and hence decrease the
strength of the 21-cm absorption.

  To compute the background of the continuum photons near the \lya
line, we assume a blackbody spectrum for the first stars. Let $E(\nu)$
be the number of photons emitted over all time per baryon and per unit
of frequency in one of the halos that collapse to form the first stars.
In a halo that forms a single, massive metal-free first star, the
spectrum should be a blackbody with temperature $T\simeq 10^5$ K, with
little dependence on the stellar mass (Schaerer 2002). As more massive
halos collapse and continue to form stars from metal-polluted gas, 
different stellar populations should form and emit more photons. For
simplicity, we shall assume
here that $E(\nu)$ remains constant, that is to say, the total number
and spectrum of photons emitted per baryon in collapsed halos is fixed
at frequencies greater than \lya (which are the only ones that affect
our result). In view of our present ignorance on the type of stars that
will form in halos of different mass as the processes of collapse of
structure in the universe, reionization and metal pollution proceed,
this assumption seems as good as any other.

  The fraction of mass bound in star-forming halos is
\begin{equation}
F(z) = \int_{M_{\rm min}(z)}^{\infty} 
m \frac{dN(m,z)}{dm} dm, 
\end{equation}
where $\frac{dN(m,z)}{dm}$ is the comoving density of halos of mass $m$,
and $M_{\rm min}(z)$ is the minimum halo mass to form the first stars at
redshift $z$. The star formation time and the main-sequence lifetime of
the first stars are short compared to the Hubble time, so the emissivity
can be approximately written as
\begin{equation}
\epsilon(\nu,z) = n_H E(\nu) \dot{F} ~.
\end{equation}
where $\dot F$ is the time derivative of the fraction of mass bound in
star-forming halos, $F$, and we have assumed a constant $E(\nu)$.

  As mentioned above, the only photons that contribute to the \lya
background are those emitted by the stars in the continuum to the blue
of the \lya line. Among these, only photons with $\nu<\nu_{\beta}$ can
be redshifted to $\nu_{\alpha}$ as continuum photons, and all photons of
greater frequency will be absorbed at the Lyman $\beta$ or higher
resonance Lyman series lines and be converted to injected \lya photons.
Here we will calculate only the intensity of the continuum \lya
background, and we will need to bear in mind that the total background
will actually be a little bit larger because of the injected photons from
higher Lyman series lines (note that the spin changes caused by these
higher Lyman series photons before they convert to \lya photons are
negligible, because of the very large number of scatterings that the \lya
photons undergo). This intensity is given by
\begin{equation}
\frac{J}{\tilde{J}_0} = E_{\alpha} \nu_{\alpha} \,
\int_z^{z_{\beta}} \frac{dF(z')}{dz'}\,
E\left[\frac{1+z'}{1+z} \nu_{\alpha}\right]/E_{\alpha}\, dz' ~,
\end{equation}
where $E_{\alpha} = E(\nu_{\alpha})$.
If the spectrum of first stars $E(\nu)$ is approximated as flat near the
\lya frequency (which is a good approximation for massive, metal-free
stars, which have surface temperature of $10^5$ K), 
then the expression is simplified to 
\begin{equation}
\frac{J}{\tilde{J}_0} = E_{\alpha} \nu_{\alpha}\, [F(z)-F(z_{\beta})] ~.
\end{equation}
We obtain an estimate for the value of $E_{\alpha}$ as follows: a
typical halo with baryonic mass of $10^5 \Msun$ forming one star with
$10^2 \Msun$ uses a fraction $10^{-3}$ of its mass to form stars. If
$\sim$ 50\% of this mass is hydrogen that is burned to helium in the
stellar interior during the star lifetime, the energy released per
baryon in the halo is $e_b = 7\, \mev\times 10^{-3}\times 0.5 = 3.5
\keV$. Assuming a blackbody spectrum with photosferic temperature $T$,
we find
\begin{equation}
E_{\alpha} \nu_{\alpha} = \frac{15}{\pi^4}
\frac{1}{\exp (h\nu_{\alpha}/kT) - 1 } \left( \frac{h\nu_{\alpha}}{kT}
\right)^4 \frac{e_b}{h\nu_{\alpha} } ~.
\end{equation}
For $T=10^5$ K, this expression gives $E_{\alpha} \nu_{\alpha} \simeq
45$. We use this value to compute the background intensity.

\begin{figure}[tbp]
\begin{center}
\includegraphics[width=0.42\textwidth]{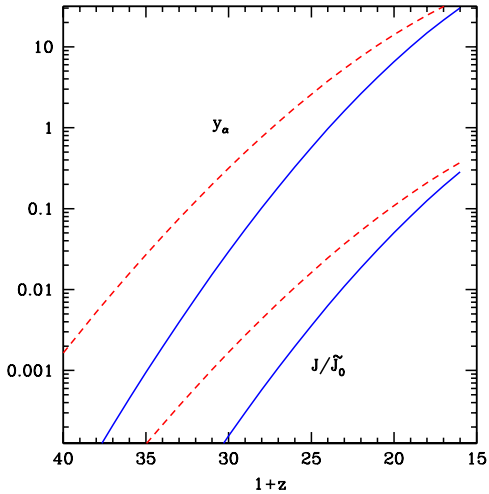}
\caption{\label{fig:jratio} Ratio $J/\tilde{J}_0$ (lower curves)
and $y_{\alpha}$ (upper curves) as a function of redshift. 
Solid curves are for the Press-Schechter mass function, dashed curves
are for the Sheth-Tormen mass function.}
\end{center}
\end{figure}

  The fraction of mass bound in halos above the minimum mass to form
stars is computed for the two halo models we use, i.e., the 
\citet{PS74} model,
\begin{equation}
F= \erfc(\sqrt{s/2}) ~,
\end{equation}
and the \citet{ST99} model,
\begin{equation}
F= 0.4 \left[1+\frac{0.4}{s^{0.2}}~ 
\erfc(0.85\sqrt{s/2})\right] ~,
\end{equation}
where $s=\delta_c^2(z)/\sigma^2(M_{min})$.
The resulting \lya intensity is plotted as a function of $z$
in Figure \ref{fig:jratio}. The corresponding $y_{\alpha}$, which gives
the strengh of the spin-kinetic coupling by \lya scattering, is also
shown in the same figure.

  The figure shows the \lya background intensity becomes important at
very high redshift. For the Sheth-Tormen model, $y_{\alpha} >1 $ at
$z<28$, while for the Press-Schechter model, this is only slightly
delayed (at $z=25$).

  It may seem surprising that the background becomes important at such
an early stage in the formation of the first generation of stars,
because as mentioned earlier the flux of local \lya photons in a \lya
sphere is actually dominated by the secondary \lya photons produced from
the ionization by X-rays, whereas the \lya background is contributed
only by the direct ultraviolet emission from the distant stars. But this
is compensated for by the large number of first stars that are already
visible from any location out to a large distance in the universe at
these redshifts. This can be seen considering, as an illustration, that
each star in the universe emits a luminosity $L_i$ in X-rays that are
absorbed at radius $ \sim r_{\alpha}$ and have their energy converted
into ``injected'' \lya photons, and a luminosity $L_c$ of ``continuum''
photons in each frequency range $\Delta\nu /\nu_{\alpha} =
H r_{\alpha}/c$, which enter the resonance \lya line as they propagate
through a distance equal to the size of the \lya sphere.
Then, the local flux of \lya photons from the star (dominated by the
injected ones) is $F_i = L_i/(4\pi r_{\alpha}^2)$, and the background
flux (determined by the continuum photons) is $F_b = n_{\alpha} L_c
r_m$, where $n_{\alpha}$ is the number density of \lya spheres with
first stars of similar luminosity, and
$r_m$ is the maximum distance out to which these first stars can be
seen. In practice, $r_m$ is determined by the extremely fast evolution
of the density of first stars (which parallels that of the background
intensity shown in Fig.~\ref{fig:jratio}), which doubles every
$\Delta z \simeq 2$. Hence, $r_m \sim cH^{-1} \Delta z/(1+z)$. Defining
the filling factor of the \lya spheres
$f=4\pi n_{\alpha} r_{\alpha}^3 /3$, we find that
\begin{equation}
\frac{F_i}{F_b} = \frac{L_i}{L_c} \frac{r_{\alpha}}{r_m} \frac{1}{3f} ~.
\end{equation}
For the first stars at $z\sim 30$ and at $r_{\alpha}=10$ kpc,
$L_i/L_c \sim 100$ (as determined from a blackbody spectrum with
$T\simeq 10^5$ K), and $r_m/r_{\alpha} \sim 300$, so the background of
continuum photons becomes equal to the locally produced injected
photons when the filling factor of the \lya spheres is about 10\%.
The fluctuation in \lya photon flux as a
function of smoothing scale was also discussed by \cite{BL04}.

  Hence, the background starts diminishing the contrast of the \lya
spheres when their filling factor becomes large. Of course, when the
filling factor becomes large the background of injected \lya photons
generated by X-rays from other stars (which we have not included here)
should also become important. The importance of other stars in the
neighborhood is greatly increased by the large biasing of the first
stars. As we can see in Fig.~\ref{fig:cumu}, we typically expect to find
several other halos capable of forming metal-free stars within the
distance of the physical size of the \lya sphere ($\sim $ 10 kpc) at $z
\lesssim 25$. This means that in practice, at $z \lesssim 25$ some of
the \lya spheres may actually be produced by ``clusters'' of first
stars, where several stars contribute to the \lya fluxes of large blobs
yielding 21-cm absorption. 

\subsection{The X-ray background}

  In addition to the soft X-rays from the stellar photosphere, the
intergalactic medium may also be heated up by a hard X-ray background,
produced for example by a population of early X-ray binaries after some
of the metal-free stars have collapsed into black holes, or by supernova
explosions and the relativistic electrons generated in supernova
remnants. The hard X-rays heat the atomic medium globally, because their
mean free path is large compared to the typical separation between \lya
spheres (as opposed to the soft X-rays from the metal-free main-sequence
stars, which do not heat the medium substantially beyond the radius of a
\lya sphere as discussed in \S 2). For local star bursts, it has been
estimated that the energy release in hard X-rays is \citep{O01}
\begin{equation}
L_X \sim 5 \times 10^{40} \erg \frac{\rm SFR}{\Msun \yr^{-1}}
\end{equation}
or about $\epsilon_X \sim 1 \keV$ per baryon that forms stars. At high
redshift the X-ray emission may increase because of the higher
temperature of the CMB which provides seed photons for Compton
scattering \citep{C03}.

  In practice, only a small fraction of the energy emitted in hard
X-rays will be used to heat the IGM. At very high frequency, the
universe is transparent to X-ray photons and so most of the emitted
energy will be redshifted rather than absorbed by an atom. At redshift
$z=30$, the universe is opaque below a photon energy of $\sim 2 \keV$.
Only a small fraction of the energy of the absorbed photons will be
converted to heat (the rest being used for collisional ionizations and
excitations). If $\eta_{eff}$ is the fraction of the emitted X-ray
energy converted to heat, and $f_*$ is the fraction of baryons in
collapsed halos with $M> M_{min}(z)$ that form stars, then the
temperature evolution of the gas is described by (neglecting other
heating and cooling mechanisms such as shock heating, Compton effect, etc.) 
\begin{equation}
\frac{dT}{dz} = -F\frac{f_* \eta_{eff} \epsilon_X}{\mu k_B}
 +\frac{2T}{1+z}.
\end{equation}

\begin{figure}[tb]
\begin{center}
\includegraphics[width=0.45\textwidth]{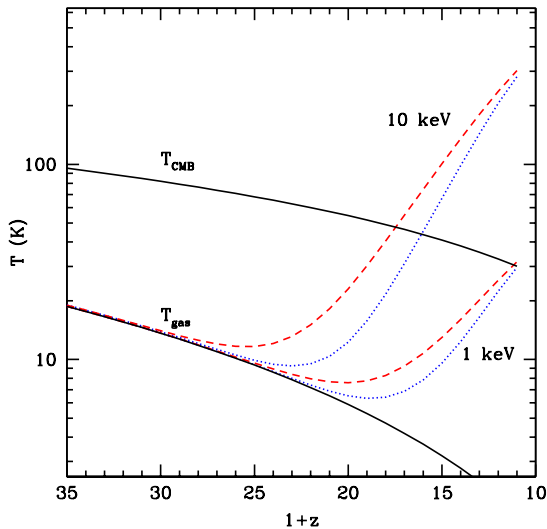}
\caption{\label{fig:Tevol} Temperature evolution of the IGM.
Solid curves show $T_{CMB}$ and the adiabatically evolved gas
temperature. Short and long-dash curves show the IGM heated by X-rays,
with the PS (short dash) and ST (long dash) model calculations. Two sets
of curves are plotted, for $\epsilon = 1 \keV$ and $10 \keV$ respectively.}
\end{center}
\end{figure}

  To see a characteristic example for the gas temperature evolution, we
integrate this equation using $f_* \eta_{eff} = 2\times 10^{-5}$ (a reasonable
value obtained for $f_* \sim 2\times 10^{-3}$ for a typical metal-free star
formed in a halo with $\sim 10^5 \Msun$ of baryons, and a fraction
$\eta_{eff}\sim 10^{-2}$ if 10\% of the hard X-ray energy is emitted
in the relevant range $0.2 - 2$ keV and 10\% of this energy is
converted into heat). We start at $z=60$ with a mean gas temperature of
66.6 K as given by the code RECFAST \citep{sss00}. The results are shown
in Fig.~\ref{fig:Tevol}. The CMB temperature and the adiabatically
evolved gas temperature are also shown. The evolution of the mean
temperature of the IGM depends sensitively on the hard X-ray emission.
For $\epsilon_X \sim 1 \keV$ per baryon, the mean
gas kinetic temperature is below that of the CMB down to $z\sim 11$, but
for $\epsilon_X \sim 10 \keV$, the gas is heated above the CMB
temperature at $z\sim 15 -17$ in our model.

\begin{figure}[tb]
\begin{center}
\includegraphics[width=0.45\textwidth]{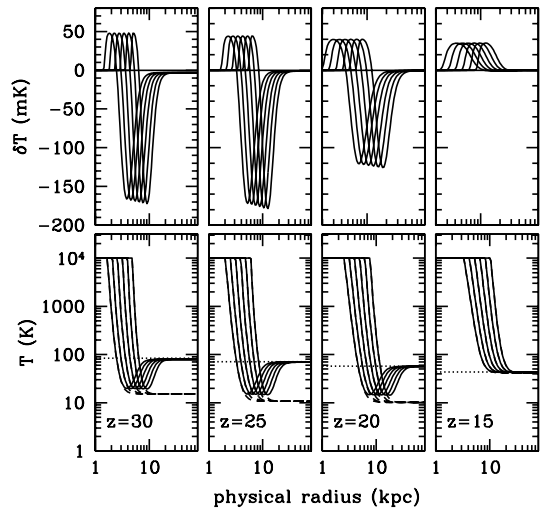}
\caption{\label{fig:profh} Brightness and temperature profiles of \lya
spheres, as in Fig.~\ref{fig:profO} except with X-ray heating of 10 keV
per baryon.}
\end{center}
\end{figure}

\begin{figure}[p]
\begin{center}
%\plotone{cmap.eps}
\includegraphics[width=0.8\textwidth]{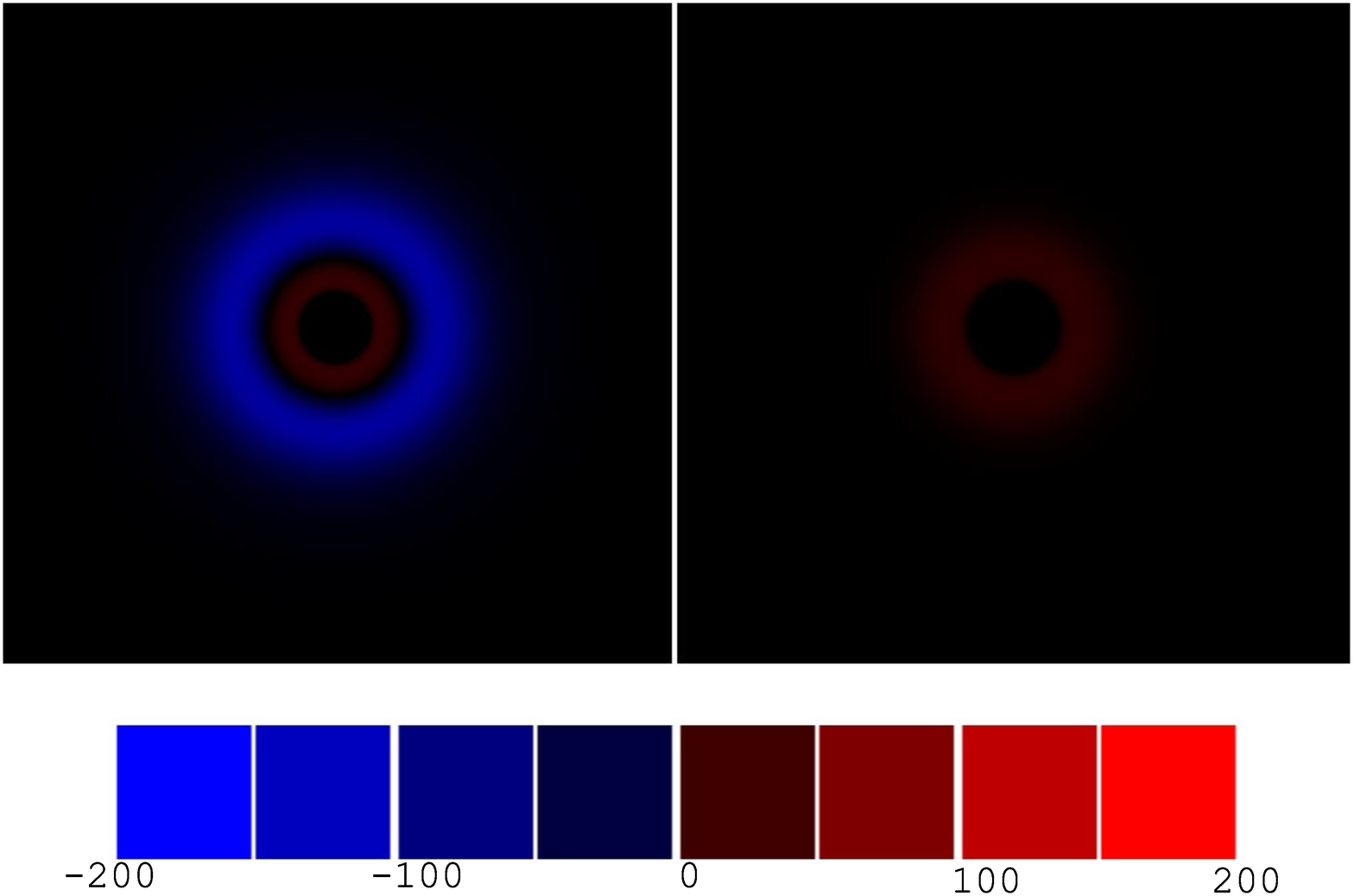}
\caption{\label{fig:cmap} A cross section map for a star of 200
  $\Msun$ and age 1.5 Myr at redshift 20(left) and 15(right). The box
  size is 40 kpc across (physical distance), and unit of temperature
  scale is mk.}
\end{center}
\end{figure}

  We plot the profiles of the \lya sphere in the presence of hard X-ray
heating in Fig.~\ref{fig:profh}, with $\epsilon_X = 10 \keV$ per baryon,
$f_* \eta_{eff} = 2 \times 10^{-5}$, and using PS halo model. At high redshift
($z=30$ to 25), when X-ray heating is still not significant, the
profiles are similar to those shown in Fig.~\ref{fig:profO}. At lower
redshifts, however, the gas is gradually heated and therefore the
21-cm absorption becomes gradually weaker. For $\epsilon = 10 \keV$,
the \lya sphere would of course appear in emission everywhere for
$z \lesssim 20$ because the gas kinetic temperature is globally heated
above the CMB temperature (see the right hand panel of
Fig~\ref{fig:cmap}, which shows a 200 $\Msun$ star at $z=15$ where the 
mean temperature of the gas is already heated above the CMB
temperature).

  It needs to be born in mind that the history of heating of the IGM is
completely dependent on the type of X-ray sources that appeared among
the first population of stars. These are likely to be X-ray binaries
made by the first stellar black holes, in stars that are formed as
close binaries. It is also possible that a population of miniquasars
made when gas can cool in the vicinity of these first black holes
emitted X-rays efficiently. The parameters we have chosen here for the
X-ray emission should be considered as no more than an illustrating
example, because we cannot predict the abundance of these first X-ray
sources.  The possibility of observing the transition
from \lya spheres with a strong absorption signal to a weaker emission
signal at lower redshift would provide a very useful tool to study the
thermal evolution of the atomic IGM and the nature of the first sources
in the universe.

Our simple estimate of the \lya and X-ray background are in general
agreement with other analytical models, e.g. \citet{S05,F06}. 
However, \citet{S05} noted that the 
the actual process of X-ray heating may be an inhomogeneous one. 
Some part of the IGM gas would be heated earlier and to higher 
temperature than other parts. With such temperature variations, 
one would observe that for the same redshift, some 
\lya spheres have only weak emission signal, while others 
have strong absorption signal. Observation of
such phenomena would reveal the temperature variations of the IGM (see
 also \citealt{PF07}).

\section{Observability}

  So far we have presented a theoretical investigation of the properties
of what we have named `` \lya spheres'', the first regions of the
universe illuminated by \lya photons from a metal-free star or a
``cluster'' of such stars (where the clusters would be caused by biasing
in the cosmological distribution of star-forming halos). Would it be 
possible to observe this object directly with the incoming 21-cm
instruments, such as 21CMA/PAST, LOFAR, MWA, and SKA? Or, if not these
instruments, can we design one which could do it? What would be the
characteristics of an instrument optimized for detecting the \lya sphere
of the first stars?

If the instrument has a response function $R(\nu,\vec{\theta})$, with
$\int d\nu d\theta^2 R(\nu,\vec{\theta})=1$, then for a pixel centered
on the \lya sphere, the resulting signal is 
\begin{equation}
\hat T=\int d\nu d\theta^2 R(\nu,\vec{\theta})
\delta T(\nu,\vec{\theta}) ~.
\end{equation}
We can model $R(\nu,\vec{\theta})$ as a direct product of a top-hat
spectral function with a bandwidth $\Delta\nu$ and a Gaussian beam of
half-beam width $\Theta$. Then,
\begin{equation}
\delta \hat T = \frac{1}{2\pi \Theta^2 \Delta\nu} 
\int_{\nu(z)-\frac{\Delta\nu}{2}}^{\nu(z)+\frac{\Delta\nu}{2}} 
d\nu \int_0^{\infty} d^2\theta
~ \delta T(\nu,\vec{\theta}) e^{-\theta^2/2\Theta^2} ~.
\end{equation}

For an interferometer array, the real space pixel noise 
is given by \citep{TMS01,SCK04}
\begin{equation}
\Delta T \sim \frac{T_{sys}}{f_{cov}\sqrt{\delta \nu t}} ~,
\end{equation}
where $T_{sys}$ is the system temperature, $\delta \nu$ is bandwidth,
$t$ the observation time, and 
\begin{equation}
f_{cov} = N_{dish} A_{dish} /A_{total}
\end{equation}
is the covering factor of the
array. Here, $A_{dish}$ is the area of an antenna, and $A_{total}$ is
the total area of the square made by the base line of the array.
The angular resolution of the array is
\begin{equation}
\Theta \sim \left(\frac{A_{dish}}{A_{total}}\right)^{1/2}
 \sim \frac{\lambda}{L} ~,
\end{equation}
where $L$ is the baseline, and $\lambda$ the redshifted wavelength.
Thus, for a fixed value of $N_{dish} A_{dish}$, the covering factor
decreases with resolution. The signal to noise ratio of the measured
radio brightness within a resolution element of the array is
\begin{equation}
\label{eq:snr}
{\rm SNR} = \sqrt{\Delta\nu t} f_{cov}
\frac{\hat{\delta T}}{T_{sys}} ~.
\end{equation}
The system temperature is
dominated by the galactic synchrotron background, which scales as
$\nu^{-2.5}$, or in terms of redshift, 
\begin{equation}
T_{sys} \simeq 2000 \K \left(\frac{1+z}{21}\right)^{2.5} ~.
\end{equation}
The signal to noise ratio at redshift $z$ is then
\begin{equation}
\label{eq:snrz}
{\rm SNR} \sim 3 f_{cov} 
\left(\frac{1+z}{21}\right)^{-2.5} 
\left(\frac{\Delta \nu \cdot t}{10 \kHz \cdot \yr}\right)^{1/2}
\left(\frac{\delta T}{10 \mK}\right) ~.
\end{equation}

\begin{figure}[tb]
\begin{center}
\includegraphics[width=0.45\textwidth]{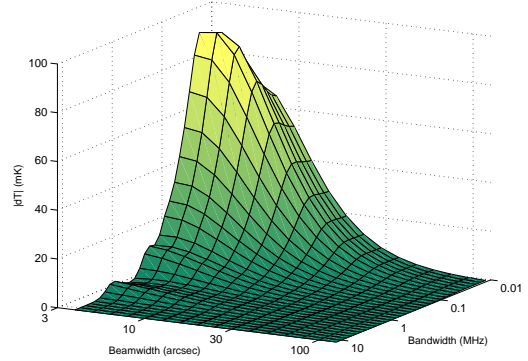}
\caption{\label{fig:sig20_400} The absolute value of the signal 
$|\delta \hat{T}|$ as a function of beamwidth $\Theta$ and 
bandwidth $\Delta \nu$, for a star at
$z=20$, with a mass of 400 $\Msun$ and an age of 1.5 Myr. }
\end{center}
\end{figure}

\begin{figure}[tb]
\begin{center}
\includegraphics[width=0.45\textwidth]{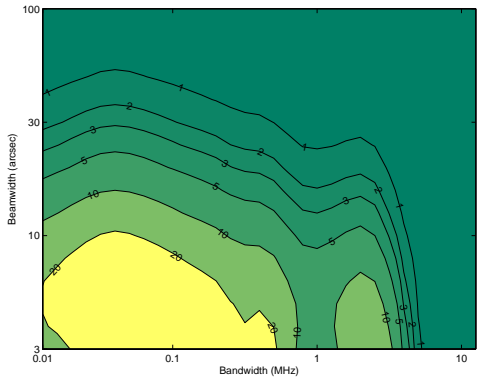}
\caption{\label{fig:sigsnr20_400} The signal to noise ratio SNR 
as a function of beamwidth $\Theta$ and bandwidth $\Delta \nu$, for a star at
  $z=20$, with a mass of 400 $\Msun$ and an age of 1.5 Myr, assuming 
$f_{cov}=1$ and $t=1 \yr$. }
\end{center}
\end{figure}

  At redshift $z$, and for a flat space geometry, a comoving distance
$dr$ is related to angular size $d\theta$ and frequency variation $d\nu$
by
\begin{equation}
d\theta = dr/r, \qquad d\nu=\nu_0 \frac{H dr}{c(1+z)^2},
\end{equation}
where $r= \int c\, dz/H$. The comoving distance is related to the
physical distance by $r=(1+z) r_p$. For the \lya spheres, the relevant
physical scale is $\sim 10 \kpc$. For the set of cosmological parameters
adopted by us, a physical distance of 10 kpc at redshift $z=20,30$
corresponds to an angular scale of 3.9, 5.3 arcsec and a frequency
interval of 7.6, 9.2 kHz, respectively.

In a resolved observation,
the maximum absorption signal strength can reach a value close to
$ \delta T \sim -200 \mK$ (Fig.~\ref{fig:profO}). However, in order to
resolve the \lya sphere, a beam of $\Theta \lesssim 4 \arcsec$ and
frequency bandwidth $\nu \lesssim 8\kHz$ are required. For observations
of the redshifted 21-cm line, the baseline for achieving an angular
resolution $\theta$ is given by 
\begin{equation}
L = 910 \left(\frac{1+z}{21}\right)
\left(\frac{\theta}{1\arcsec}\right)^{-1} \km ~.
\end{equation}
For a beamwidth and bandwidth similar to the size of the \lya sphere,
one obtains the best signal-to-noise ratio for a single brightness
measurement.
Assuming that the center of the beam and frequency bin is centered on
the star, we plot the observational signal $|\delta\hat{T}|$ of 
the \lya sphere made by a $400 \Msun$ star at $z=20$ of age
$t=1.5\Myr$, as a function of the beamwidth $\Theta$ and bandwidth 
$\Delta\nu$ in Fig.~\ref{fig:sig20_400}. 
For this star, the maximum emission brightness occurs at 6.9 kpc away
from the star, and the maximum absorption is located at 15 kpc
($\delta T=-190\mK$), but $|\delta T| > 5\mK$ up to 33 kpc.
The signal decreases more rapidly with $\Theta$
than $\Delta\nu$, which is not unexpected given that the included volume is 
$\propto \Theta^2 \Delta\nu$. There are some wiggles of $\delta T$ in 
the $\Delta\nu$ direction, probably because as we vary the size of
the frequency window, we are sampling different structure of the \lya sphere.

We plot the corresponding signal to noise ratio (SNR; see eq.\
[\ref{eq:snrz}]) in Figure \ref{fig:sigsnr20_400}, assuming
$f_{cov}=1$ and $t=1 \yr$. 
Again, we ignore the second peak at very large bandwidth.  
To obtain a high confidence detection with ${\rm SNR} > 5 (10, 20)$ 
under this circumstance, one could choose an optimal bandwidth of 30
kHz, and the optimal beamwidth (to obtain the highest SNR) is
20 (14, 10) arcsec, corresponding to a baseline of 45 (65, 91) km
.
The area corresponding to this baseline would have to be covered with
dipole receivers in order to achieve $f_{cov} \sim 1$. 
The requirements become more stringent with increasing redshift, owing
to the increasing brightness of the synchrotron foreground, and as we
have seen in \S 4 it may be necessary to go up to $z\sim 30$ to see a
pristine \lya sphere with a signal that is not yet weakened by a global
\lya background and an IGM heated by hard X-rays.

\section{Conclusion}

  We have investigated the 21-cm signature of the first stars. The
coupling of the spin and kinetic temperature induced by the \lya
photons generated by a metal-free star should generate a
``\lya sphere'', a region that can be detected by a 21-cm absorption
signal when the IGM has not been heated by X-rays prior to the formation
of the star. The 21-cm absorption signal can reach a strength of 200 mK
when the IGM temperature far from the star is at its minimum value
reached by the adiabatic cooling, and becomes weaker as the IGM is
heated. Most of the \lya photons that couple the spin and kinetic
temperatures are generated by collisional excitations of the high-energy
electrons that are produced by soft X-rays emitted from the photospheres
of the hot metal-free stars, rather than the direct photospheric
emission of the star. These soft X-rays are not able to travel very far
from the star and hence tend to heat the IGM only locally around each
metal-free star that forms, but the IGM may be heated globally by other
sources of hard X-rays, such as X-ray binaries formed by the black
holes made by the first stars.

  The temperature and 21-cm absorption profile of a \lya sphere can be
calculated as a function of the spectrum emitted by the star and its
age in the idealized case of a uniform medium in Hubble expansion around
the star. However, owing to the highly biased distribution of the first
stars, the region where a first star has formed may often be highly
irregular, containing non-linear structures on scales larger than the
halo hosting the star. The first stars should also be highly clustered
because of their biased distribution, so the high-redshift regions
producing 21-cm absorption may often be produced by several first stars
that are illuminating and heating a larger region with their X-rays and
\lya photons. This may make these regions larger and therefore easier to
detect. 

  The detection of these earliest structures in the universe formed by
the first stars is still very challenging, mainly because of the high
redshift at which they are likely to be present and the intense
synchrotron foreground that must be subtracted to measure any CMB
fluctuations. Filled aperture of at least $\sim 50$ km will be
required for detection \footnote{After the preprint of this paper
was circulated, it was shown in \cite{LZC07}
that the angular size of the \lya sphere can be magnified by 
gravitational lensing. For some strongly lensed system, the
baseline could be reduced to about 10 km.}. However, detecting 
these regions of \lya
absorption will represent a milestone for cosmology: they will constitute a
clear indication that the first stars formed in the universe have been
found, and allow us to measure many of their properties such as their
mass distribution and spatial correlation.

\acknowledgments
We thank Chris Hirata, Steven Furlanetto, Ue-Li Pen, Pengjie Zhang,
and the anonymous referee for discussions and suggestions. 
XC acknowledges the ICRA and the 
Caltech TAPIR group for their hospitality during his visit. XC
is supported by the NSFC under the Distinguished Young Scholar 
grant No. 10525314, the Key Project grant No. 10533010, by the CAS
under grant KJCX3-SYW-N2, and by the
Ministry of Science of Technology under the National Basic Science
Program (Project 973) under grant No. 2007CB815401.

\begin{thebibliography}{}

\bibitem[Abel, Bryan, Norman(2000)]{ABN00} Abel, T., Bryan, G. \&
  Norman, M. L., 2000, \apj, 540, 39.

\bibitem[Abel, Bryan, \& Norman(2002)]{ABN02} Abel, T.,
Bryan, G.~L., \& Norman, M.~L.\ 2002, Science, 295, 93

\bibitem[Allison \& Dalgarno(1969)]{AD69} Allison, A. C. \& Dalgarno,
A., 1969, \apj, 158, 423.

\bibitem[Barkana \& Loeb(2001)]{BL01} Barkana, R., Loeb, A., 2001, 
\physrep, 349, 125.

\bibitem[Barkana \& Loeb(2005a)]{BL04} Barkana, R., Loeb, A.,
  2005, \apj, 626, 1.

\bibitem[Barkana \& Loeb(2005b)]{BL05} Barkana, R., Loeb, A., 2005,
  \mnras, 363, L36.

\bibitem[Bharadwaj \& Ali(2004)]{BA04} Bharadwaj, S., Ali, S. S.,2004, 
  \mnras, 142. 

\bibitem[Bromm, Coppi, \& Larson(1999)]{BCL99} Bromm, V., Coppi, P.S.,
  Larson, R. B., 1999, \apj, 527, 5.

\bibitem[Bromm, Kudritzki \& Loeb(2001)]{BKL01} Bromm, V., 
Kudritzki, R. P., \& Loeb, A., 2001, \apj, 552, 464.

\bibitem[Bromm \& Larson(2004)]{BL04rev} Bromm, V., \& Larson, R. B.,
  2004, \araa, 42, 79.  

\bibitem[Cen(2003)]{C03} Cen, R., 2003, \apj, 591, 12.

\bibitem[Cen(2006)]{C06} Cen, R., 2006, \apj, 648, 47.

\bibitem[Chen \& Miralda-Escud\'e(2004)]{CM04} Chen, X. \&
Miralda-Escud\'e, J., 2004, \apj, 602, 1.

\bibitem[Chen \& Kamionkowski(2004)]{CK04} Chen, X., \& Kamionkowski, M., 
2004, \prd, 70, 043502.

\bibitem[Ciardi \& Ferrara(2005)]{CF05} Ciardi, B. \& Ferrara, A.,
  2005, \ssr, 116, 625.

\bibitem[Ciardi \& Madau(2003)]{CM03} Ciardi, B. \& Madau, P.,2003, \apj,596,1.

\bibitem[Cohn \& White(2007)]{CW07} Cohn, J. D., \& White, M., 2007,
  arXiv:0706.0208.

\bibitem[Engeln-Mullges \& Uhlig(1996)]{EU96} Engeln-Mullges, G. \&
  Uhlig, F., 1996, {\it Numerical Algorithms with C}, Springer, 1996.

\bibitem[Field(1959)]{F59} Field, G. B., 1959, \apj, 129, 551.

\bibitem[Furlanetto(2006)]{F06} Furlanetto, S., 2006, \mnras, 371, 867.

\bibitem[Furlanetto \& Furlanetto(2007a)]{FF07a} Furlanetto, S. \&
  Furlanetto, M., 2007a, \mnras, 374, 547.

\bibitem[Furlanetto \& Furlanetto(2007b)]{FF07b}  Furlanetto, S. \&
  Furlanetto, M., 2007b, \mnras, 379,130.

\bibitem[Furlanetto, Oh \& Briggs(2006)]{FOB06} Furlanetto, S.,
Oh,  S. P., \& Briggs, F. H., 2006, \physrep, 433, 181.

\bibitem[Furlanetto, Oh \& Pierapoli(2006)]{FOP06} Furlanetto, S.,
  Oh, S. P., \& Pierpaoli, E., E., 2006, \prd 74, 103502.


\bibitem[Furlantto \& Pritchard(2006)]{FP06} Furlanetto, S. \&
  Pritchard, J. R., 2006, \mnras, 372, 1093.

\bibitem[Furlanetto, Sokasian, \& Hernquist(2004)]{FSH04} Furlanetto, S.,
  Sokasian, A., Hernquist, L., 2004, \mnras, 347, 187.

\bibitem[Hirata(2006)]{H06} Hirata, C., 2006, \mnras, 367, 259.

\bibitem[Hummer \& Seaton(1963)]{HS63} Hummer, D. G., Seaton, M. J.,
  1963, \mnras 125, 437.

\bibitem[Hummer \& Storey(1998)]{HS98} Hummer, D. G., Storey, P. J.,
   1998, \mnras, 297, 1073.

\bibitem[Iliev et al(2002)]{I02} Iliev, I. T., Shapiro, P. R.,
  Ferrara, A., Martel, H., 2002, \apjl, 572, L123.

\bibitem[Iliev et al.(2006)]{I06} Iliev, I. T., Mellema, G., Pen,
  U.-L., Merz, H., Shapiro, P.R., Alvarez, M. A., 2006, \mnras, 369, 1625. 

\bibitem[Jang-Condell \& Hernquist(2001)]{JH01} Jang-Condell, H.,
  Hernquist, L., 2001, \apj, 548, 68.

\bibitem[Jenkins et al.(2001)]{J01} Jenkins, A. et al., 2001, \mnras,
  321, 372. 

\bibitem[Kohler et al(2005)]{K05} Kohler, K., Gnedin, N. Y.,
Miralda-Escud\'e, J., \& Shaver, P. 2005, \apj, 633, 552.

\bibitem[Kuhlen et al(2006)]{K06} Kuhlen, M., Madau, P., \&
 Montgomery, R. 2006, \apj, 637, 1

\bibitem[Li, Zhang, \& Chen(2007)]{LZC07} Li, G., Zhang, P., Chen, X., 
2007, \apj, 666, 45.

\bibitem[Loeb \& Rybicki(1999)]{LR99} Loeb, A. \& Rybicki, G. B.,
  1999, \apjl 524, L527.

\bibitem[Loeb \& Zaldarriaga(2004)]{LZ04} Loeb, A.\& Zaldarriaga, M., 2004,
\prl, 92, 1301.

\bibitem[Lukic et al.(2007)]{L07} Lukic, Z., Heitmann, K., Habib, S.,
  Bashinsky S., Ricker, P. M., 2007, arXiv:astro-ph/0702360

\bibitem[Madau, Meiksin, \& Rees(1997)]{MMR97} Madau, P., Meiksin, A.,
  \& Rees, M. J., 1997, \apj, 475, 492.


\bibitem[Miralda-Escud\'e(2003)]{M03} Miralda-Escud\'e, J., 2003, 
{\it Science}, 300, 1904.

\bibitem[Mo \& White(1996)]{MW96} Mo, H.J. \& White, S.D.M., 1996,
  \mnras, 282, 347.

\bibitem[Mo \& White(2002)]{MW02} Mo, H.J. \& White, S.D.M., 2002,
  \mnras, 336, 112.

\bibitem[Naoz \& Barkana(2005)]{NB05} Naoz, S., Barkana, R., 2005, \mnras,
  362, 1047.

\bibitem[Oh(2001)]{O01} Oh, S.P., \apj 553, 499.

\bibitem[Omukai \& Palla(2003)]{OP03}Omukai, K., \& Palla, F. 2004,
 \apj, 589, 677.

\bibitem[Press \& Schechter(1974)]{PS74}Press, W., \& Schechter, P.,
  1974, \apj, 187, 425.

\bibitem[Pritchard \& Furlanetto(2006)]{PF06} Pritchard, J. R., \&
Furlanetto, S., 2006, \mnras, 367, 1057.

\bibitem[Pritchard \& Furlanetto(2007)]{PF07} Pritchard, J. R., \&
Furlanetto, S., 2007, \mnras, 376, 1680.


\bibitem[Reed et al.(2003)]{R03} Reed, D. et al, 2003, \mnras, 346, 565.

\bibitem[Reed et al.(2005)]{R05} Reed, D. S. et al., 2005, \mnras,
  363, 393.


\bibitem[Ripamonti, Mapelli, Ferrara(2007)]{RMF07} Ripamonti, E.,
  Mapelli, \&M., Ferrara, A., 2007, \mnras, 375, 1399.

\bibitem[Santos, Cooray \& Knox(2004)]{SCK04} Santos, M.G., Cooray,
  A., \& Knox, L., \apj, 625, 575.

\bibitem[Seager, Sasselov \& Scott(2000)]{sss00} Seager, S., 
Sasselov, D., \& Scott, D. 2000, \apjs, 128,
407.

\bibitem[Sethi(2005)]{S05} Sethi, S. K., 2005, \mnras 363, 818.

\bibitem[Sheth \& Tormen(1999)]{ST99} Sheth, R. K. \& Tormen, G., 1999,
\mnras, 308,119. 

\bibitem[Shull \& van Steenberg(1981)]{SvS85} Shull, M. \& van
  Steenberg, 1985, \apj 298, 268.

\bibitem[Storey \& Hummer(1995)]{SH95} Storey, P. J., \& Hummer,
  D. G., 1995, \mnras, 272, 41.

\bibitem[Thompson, Moran, \& Swenson(2001)]{TMS01} Thompson, A.R.,
  Moran, J.M., Swenson, G.W. Jr., 2001, {\it Interferometry and synthesis in
  radio astronomy}, 2nd.ed., Wiley, New York.

\bibitem[Tozzi et al.(2000)]{TMMR00} Tozzi, P., Madau, P. Meiksin, A.,
\& Rees, M. J., 2000, \apj, 528, 597.

\bibitem[Valdes et al al.(2007)]{VFMP07} Valdes, M., Ferrara, A.,
  Mapelli, M., Ripamonti, E., 2007, \mnras, 377, 245.

\bibitem[Verner et al.(1996)]{V96} Verner, D.A., Ferland, G.J.,
  Korista, K.T., Yakovlev, D.G., 1996, \apj, 465, 487.

\bibitem[Warren et al.(2006)]{W06} Warren, M. S., Abazajian, K., Holz,
  D., Teodoro, L., 2006, \apj, 646,881.

\bibitem[Whalen, Abel, \& Norman(2004)]{WAN04} Whalen, D., Abel, T., Norman,
  M. L., 2004, \apj, 610, 14.

\bibitem[Woosley, Heger, \& Weaver(2002)]{WHW02} Woosley, S. E.,
  Heger, A., Weaver, T. A., 2002, RMP, 74, 1015. 

\bibitem[Wouthuysen(1952)]{W52} Wouthuysen, S. A., 1952, \aj, 57, 31.

\bibitem[Yoshida et al.(2003)]{Y03} Yoshida, N., Abel, T., Hernquist,
  L., Sugiyama, N., 2003, \apj, 592, 645.

\bibitem[Zahn et al.(2006)]{Z06} Zahn, O., Lidz, A., McQuinn, M.,
  Dutta, S., Hernquist, L., Zaldarriaga, M., Furlanetto, S. R., 
\apj, 654, 12.

%\bibitem[Zaldarriaga, Furlanetto \& Hernquist(2004)]{ZFH04}
%  Zaldarriaga, M., Furlanetto, S.R.,  \& Hernquist, L., 2004, \apj,
%  608, 622.

\bibitem[Zygelman (2005)]{Z05} Zygelman, B., 2005, \apj, 622, 1356.



\end{thebibliography}
\end{document}